\documentclass[useAMS,usenatbib]{mnras}
\usepackage[T1]{fontenc}
\usepackage{ae,aecompl}
\usepackage{pdflscape}
\usepackage{graphicx}
\usepackage{longtable}
\usepackage{lscape}
\usepackage{amssymb}
\usepackage{endnotes} 
\usepackage{footnote}
\usepackage{subfigure}
\usepackage{amsmath}
\usepackage{txfonts}
\usepackage{natbib}
\usepackage{url}
\usepackage{breakurl}
\usepackage{helvet}
\newcommand{\hcc}{SN~2017hcc}

\title[SN~2017hcc: Observational behaviour]{On the observational behaviour of the highly polarized Type IIn supernova SN~2017hcc}
\author[B. Kumar et al.]
{Brajesh Kumar$^{1}$\thanks{E-mail: brajesh.kumar@iiap.res.in, brajesharies@gmail.com},
{Chakali Eswaraiah}$^{1,2}$,
{Avinash Singh}$^{1,3}$,
{D. K. Sahu}$^{1}$,
{G. C. Anupama}$^{1}$,
\newauthor
{K. S. Kawabata}$^{4}$,
{Masayuki Yamanaka}$^{4,5}$,
{Ikki Otsubo}$^{7}$,
{S. B. Pandey}$^{6}$,
{Tatsuya Nakaoka}$^{7}$,
\newauthor
{Miho Kawabata}$^{7}$,
{Amar Aryan}$^{6}$ and
{Hiroshi Akitaya}$^{4}$\\\\
$^{1}$Indian Institute of Astrophysics, II Block, Koramangala, Bengaluru 560 034, India \\
$^{2}$National Astronomical Observatories, Chinese Academy of Sciences, Datun Road, Chaoyang District, Beijing 100101, People's Republic of China \\
$^{3}$Joint Astronomy Programme, Department of Physics, Indian Institute of Science, Bengaluru 560 012, India \\
$^{4}$Hiroshima Astrophysical Science Center, Hiroshima University, 1-3-1 Kagamiyama, Higashi-Hiroshima, Hiroshima 739-8526, Japan \\
$^{5}$Okayama Observatory, Graduate School of Science, Kyoto University, 3037-5 Honjo, Kamogata, Asakuchi, Okayama, 719-0232, Japan \\
$^{6}$Aryabhatta Research Institute of Observational Sciences, Manora Peak, Nainital 263 001, India \\
$^{7}$Department of Physical Science, Hiroshima University, Kagamiyama 1-3-1, Higashi-Hiroshima 739-8526, Japan
}

\begin{document}

\label{firstpage}
\date{Accepted 2019 July 03. Received 2019 June 06; in original form 2019 January 24}
\pagerange{\pageref{firstpage}--\pageref{lastpage}} \pubyear{2019}
\maketitle

\begin{abstract}
We present the results based on photometric (\textit{Swift} UVOT), broad-band polarimetric ($V$ and $R$-band) 
and optical spectroscopic observations of the Type IIn supernova (SN) 2017hcc. Our study is supplemented with 
spectropolarimetric data available in literature for this event. The post-peak light curve evolution is slow 
($\sim$\,0.2 mag 100 d$^{-1}$ in $b$-band). The spectrum of $\sim$\,+27 d shows a blue continuum with narrow 
emission lines, typical of a Type IIn SN. Archival polarization data along with the \textit{Gaia} DR2 distances 
have been utilized to evaluate the interstellar polarization (ISP) towards the SN direction which is found to
be $P_{ISP}$ = 0.17\,$\pm$\,0.02 per cent and $\theta_{ISP}$ = 140$\degr$\,$\pm$\,3$\degr$. To extract the 
intrinsic polarization of SN~2017hcc, both the observed and the literature polarization measurements were 
corrected for ISP. We noticed a significant decline of $\sim$\,3.5 per cent ($V$-band) in the intrinsic level 
of polarization spanning a period of $\sim$\,2 months. In contrast, the intrinsic polarization angles remain 
nearly constant at all epochs. Our study indicates a substantial variation in the degree of asymmetry in 
either the ejecta and/or the surrounding medium of SN~2017hcc. We also estimate a mass-loss rate of 
$\dot M$ = 0.12 M$_{\sun}$ yr$^{-1}$ (for $v_w$ = 20 km s$^{-1}$) which suggests that the progenitor of 
SN~2017hcc is most likely a Luminous Blue Variable.
\end{abstract}


\section{Introduction}\label{intro}

Supernovae (SNe), of Type IIn are hydrogen rich core collapse supernovae (CCSNe\footnote{These events 
characterize the explosion of massive stars ($>$8 M$_{\sun}$) \citep[for a review, see][]{1997ARA&A..35..309F}.}). 
They are primarily identified by their strong narrow and intermediate-width emission components of H, 
superimposed on a relatively smoother blue continuum \citep{1990MNRAS.244..269S}. These narrow emissions 
are believed to arise due to an interaction between the slow moving circumstellar material (CSM) and the faster 
moving SN ejecta \citep{1994ApJ...420..268C,1994MNRAS.268..173C,2001MNRAS.326.1448C}. 
Multi-wavelength studies suggest that they cover a broad range of observational properties 
\citep{2012ApJ...744...10K,2013AJ....146....2F}, pointing to the existence of sub-categories 
of Type IIn SNe \citep[see][]{2013A&A...555A..10T,2014ApJ...788..154O,2017hsn..book..403S}.
The observed diversity depends upon various factors such as progenitor channels 
\citep{2002ApJ...572..350F,2007ApJ...656..372G,2011MNRAS.412.1639D,2013ApJ...767....1P}, 
mass-loss history \citep{2014ARA&A..52..487S}, explosion mechanisms \citep{2014A&A...569A..57M},
and environments of explosion sites \citep{2014MNRAS.441.2230H,2015A&A...580A.131T}.

The fractional population of Type IIn SNe is relatively small among the volume limited survey of
CCSNe \citep[$\sim$\,10 per cent,][]{2011MNRAS.412.1473L}. Nonetheless, considering their peculiarities, 
the light curve and spectral evolution of several events have been studied in detail.  
The polarimetric (broad-band polarimetry and spectropolarimetry) investigations of Type IIn SNe in the 
past have suggested relatively high degree of polarization in comparison to other CCSNe. In addition, 
they do not exhibit significant variation in the polarization angle
\citep[c.f.][and references therein]{2008ARA&A..46..433W,2017suex.book.....B}.
This is attributed to the intrinsic aspherical SN explosion and/or asphericity in the CSM. 
The degree of polarization varies during different phases of the supernova evolution [for example, SN~1994Y
\citep{1996ApJ...467..435W}, SN~1997eg \citep{2008ApJ...688.1186H}, SN~1998S \citep{2000ApJ...536..239L,
2001ApJ...550.1030W}, SN~2010jl \citep{2011A&A...527L...6P} and SN~2012ab \citep{2018MNRAS.475.1104B}].
While spectropolarimetry serves as a direct geometrical and dynamical probe of
chemical constituents and geometry of the SN ejecta \citep[see, e.g.][]{1982ApJ...263..902S}, 
the broad-band polarization parameters, on the other hand, provide a coarse but important picture of SN, 
especially when it is fainter.

Emergence of $\sim$\,1 per cent intrinsic polarization may imply around 20 per cent asymmetry 
\citep{1991A&A...246..481H,2005ASPC..342..330L}. In the case of Type IIn SNe, the theoretical 
interpretation of the observed polarization is not straightforward because of the difficulties 
in disentangling the polarization contributors. 
It is suggested that geometry of the SN ejecta along with the geometry of the CSM are to be accounted 
\citep{2001MNRAS.326.1448C}. There is an increasing need for additional polarimetric 
data of SNe in order to enhance our current understanding about the structure of SN explosion/ejecta. 
Multi-epoch broad-band polarization and spectroscopic observations of SN~2017hcc are presented 
here with an investigation of its observational properties.

SN~2017hcc (ATLAS17lsn) was discovered\footnote{\url{https://wis-tns.weizmann.ac.il/object/2017hcc/discovery-cert}}
($\sim$\,17.4 mag, $o$-band) in an anonymous galaxy on 2017 October 2.4 (JD~2458028.9) by Asteroid Terrestrial-impact 
Last Alert System \citep[ATLAS,][]{2011PASP..123...58T}. It was located 4.3 arcsec east and 0.5 arcsec south of the 
center of the host galaxy, with coordinates $\alpha = 00^{\rm h} 03^{\rm m} 50\fs6$, ~$\delta =-11\degr 28\arcmin 28\farcs8$.
The transient was not detected on 2017 September 30.4 (JD~2458026.9) at a limiting magnitude of 19.04.
This indicates that SN was discovered within a few days after the explosion. In the present study, we adopted a 
mean of the earliest detection and the last non-detection epochs as the date of explosion for SN~2017hcc i.e. 2017 
October 01.4 (JD~2458027.9). It was classified\footnote{\url{https://wis-tns.weizmann.ac.il/object/2017hcc/classification-cert}}
as a young Type IIn SN based on a low resolution optical spectrum obtained on 2017 October 07 with FLOYDS spectrograph 
at Siding Spring Observatory. In subsequent observations, non-detection of X-ray and radio was reported by 
\citet{2017ATel10936....1C} and \citet{2017ATel11015....1N}, respectively.

Spectropolarimetric observations of SN~2017hcc obtained on 2017 October 30 by \citet{2017ATel10911....1M} indicated 
an integrated $V$-band continuum polarization of 4.84 per cent with a polarization angle of $\sim$\,97$^\circ$.
They also noticed a weak wavelength dependence in the polarization measurement and a strong depolarization in the 
cores of the narrow emission features of H$\alpha$ and H$\beta$. \citet{2017RNAAS...1a..28P} reported early phase
results of \hcc\ and found it to be one of the brightest Type IIn events known (M$_{V} \sim$\,--20.7 mag and peak
L$_{bol}$ $\sim$\,1.3 $\times$ 10$^{44}$ erg s$^{-1}$, for a distance of about 73 Mpc).

\begin{figure}
\centering
\includegraphics[width=\columnwidth]{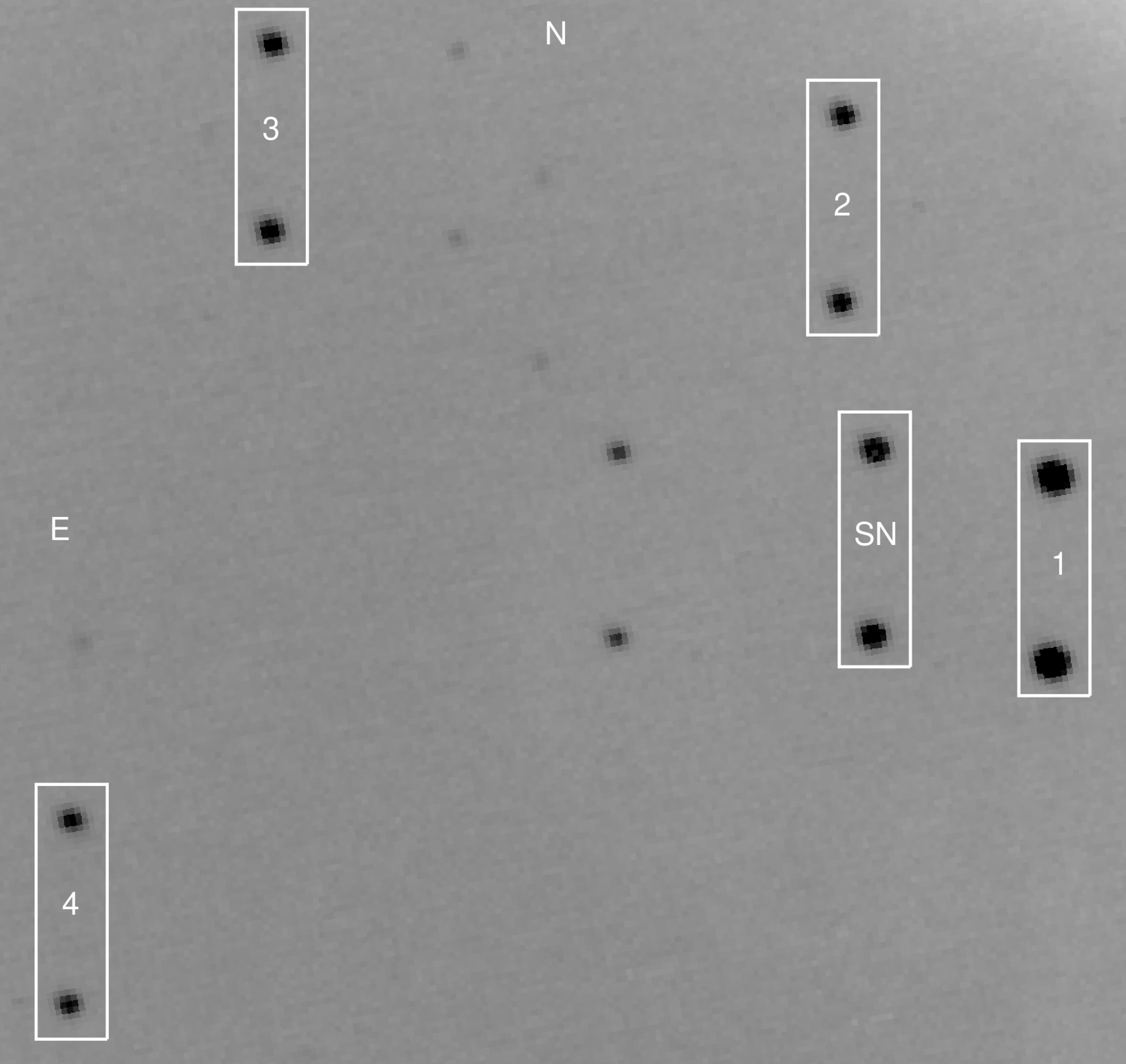}
\caption{SN~2017hcc and field stars (1\,--\,4). A zoomed AIMPOL $V$-band image (about 5 arcmin $\times$ 5 arcmin)
obtained with the ST on 2017 December 25 is shown. Each object has two images (i.e. ordinary and extraordinary)
and are marked inside rectangular boxes. North and East directions are indicated.}
\label{fig_1}
\end{figure}

\section{Observations and data reduction}\label{obs}

The $V$-band ($\lambda_{V_{eff}}$ = 0.53 $\mu$m) polarimetric data of SN~2017hcc was obtained on 2017 
December 25. These observations were acquired using the ARIES Imaging Polarimeter \citep*[AIMPOL,][]
{2004BASI...32..159R} mounted at the Cassegrain focus of the 1.04-m Sampurnanand telescope (ST) at 
Manora Peak, Nainital. This observing facility is operated by the Aryabhatta Research Institute of 
Observational sciences (ARIES), India. A half-wave plate (HWP) modulator, a Wollaston prism beam-splitter
and a liquid nitrogen cooled Tektronix 1k $\times$ 1k pixel CCD camera are integrated in the AIMPOL 
instrument. The plate scale, readout noise and gain of the CCD are 1.73 arcsec pixel$^{-1}$, 7.0 $e^{-}$ 
and 11.98 $e^{-}$/ADU, respectively. The central 370 pixel $\times$ 370 pixel region of the CCD, used for 
observation, covers $\sim$\,8 arcmin in diameter in the sky. Fig.~\ref{fig_1} displays a section 
(approximately 4 arcmin $\times$ 4 arcmin) of the full image captured with the AIMPOL using ST. AIMPOL 
has been effectively used for polarimetric observations of SNe, e.g., SN~2012aw 
\citep{2014MNRAS.442....2K}, SN~2013ej \citep{2016MNRAS.456.3157K}, SN~2013hj and SN~2014G 
\citep{2016MNRAS.455.2712B}, SN~2014J \citep{2016MNRAS.457.1000S} and SN~2016B \citep{Dastidar} etc. 

Five frames were secured at each position angle of the HWP (i.e. 0$\degr$.0, 22$\degr$.5, 45$\degr$.0 and 67$\degr$.5).
The individual exposure time for each frame was 200 seconds. To obtain a good signal-to-noise ratio (SNR), all 
pre-processed images at a given position angle were aligned and subsequently combined. The ordinary and extraordinary 
fluxes were extracted by performing standard aperture photometry using {\small IRAF}\footnote{{\small IRAF} is 
the Image Reduction and Analysis Facility distributed by the National Optical Astronomy Observatories, which are 
operated by the Association of Universities for Research in Astronomy, Inc., under cooperative agreement with the 
National Science Foundation.} software. Multiple apertures (2 to 8 pixels) were chosen to estimate Stokes parameters 
at each aperture and finally $P$ (degree of polarization) and $\theta$ (polarization angle) were computed. Detailed 
description of the various procedure and the requisite equations can be found in \citet{1998A&AS..128..369R,
2004BASI...32..159R,2014MNRAS.442....2K,2016MNRAS.456.3157K}. The considered $P$ and $\theta$ values are for the 
aperture that best fit with minimum chi-square.

The offset polarization angle was estimated by observing two polarized standard stars on the same night. For this 
purpose, HD~25443 and HD~236633 stars were selected from \citet*{1992AJ....104.1563S}. The results listed in 
Table~\ref{stand_log} are in good agreement with the standard values. Consequently, an average of offset values 
(i.e. 10.1$\degr$) was applied to the SN. The instrumental polarization of AIMPOL is generally found to be small, 
at $\sim$\,0.1 per cent \citep[see, e.g.][and references therein]{2007MNRAS.378..881M,2009MNRAS.396.1004P,
2013MNRAS.436.3500D,2012MNRAS.419.2587E,2013A&A...556A..65E,2013MNRAS.430.2154P,2015A&A...573A..34S,2016A&A...588A..45N,
2017ApJ...835..275R,2017ApJ...849..157W,2018AJ....156..115L}. This was verified by the observations 
of the unpolarized standard star HD~12021 \citep*{1992AJ....104.1563S}. 
The observed $P$ and $\theta$ values of SN ~2017hcc after correcting for the offset polarization angle and 
instrumental polarization are provided in Table~\ref{tab_log}. 

\begin{table}
\centering
\caption{Results of observed polarized and unpolarized standard stars taken from \citet*{1992AJ....104.1563S}.
\label{stand_log}}
\begin{tabular}{lcll}
\hline \hline
Standard star              & Band                   &  $P \pm \sigma_{P}$  & $\theta \pm \sigma_{\theta}$  \\
  (ID)                     &                        &  (per cent)          &            ($^\circ)$         \\
\hline
BD$+$59$^{\circ}$389 (Pol)$^{\dagger}$ & $R$        &                      &        \\
Standard values            & & 6.43 $\pm$ 0.02      & 98.14 $\pm$ 0.10              \\
Observed values            & &6.49 $\pm$ 0.02       & 98.40 $\pm$ 0.06              \\
\\
BD$+$64$^{\circ}$106 (Pol)$^{\dagger}$ & $R$        &                      &        \\
Standard values            & &5.15 $\pm$ 0.10       & 96.74 $\pm$ 0.54              \\
Observed values            & &5.35 $\pm$ 0.02       & 97.70 $\pm$ 0.17              \\
\\
HD~25443 (Pol)$^{*}$       & $V$                                                    \\
Standard values            & &5.13 $\pm$ 0.06       & 134.23 $\pm$ 0.34             \\
Observed values            & &5.50 $\pm$ 0.10       & 144.60 $\pm$ 1.00             \\
\\
HD~236633 (Pol)$^{*}$      & $V$                    &               &               \\
Standard values            & &5.49 $\pm$ 0.02       &  93.76 $\pm$ 0.08             \\
Observed values            & &5.40 $\pm$ 0.10       & 103.5  $\pm$ 2.00             \\
\hline
HD~12021 (Unpol)$^{*}$     & $V$                    &  &                            \\
Standard values            & &0.08 $\pm$ 0.02       & 160.1    ~~~~ --              \\
Observed values            & &0.12 $\pm$ 0.10       & 163.8  $\pm$ 3.50             \\
\hline
\end{tabular}\\
$^{\dagger}$ HONIR,
$^{*}$ AIMPOL \\
\end{table}

\begin{table*}
\centering
\caption{The \textit{Swift}-UVOT photometric magnitudes of SN~2017hcc (in Vega-system).}
\label{swift_mag}
\begin{tabular}{ccccccccc}
\hline \hline
Date         & JD       & Phase$^{a}$   &  $u$               &  $b$               &  $v$               &  $uvm2$            &   $uvw1$           &  $uvw2$             \\
(yyyy-mm-dd) & 2458000+ &               &  (mag)             & (mag)              &  (mag)             &  (mag)             &  (mag)             &  (mag)              \\
\hline
2017-10-28   & 54.9     &  +\,0.4       & 12.72\,$\pm$\,0.03 & 13.98\,$\pm$\,0.03 & 13.92\,$\pm$\,0.03 & 12.27\,$\pm$\,0.03 & 12.33\,$\pm$\,0.02 & 12.43\,$\pm$\,0.02  \\
2017-10-30   & 56.6     &  +\,2.1       & 12.70\,$\pm$\,0.03 & 13.92\,$\pm$\,0.03 & --                 & --                 & 12.33\,$\pm$\,0.02 & 12.47\,$\pm$\,0.03  \\
2017-11-03   & 61.3     &  +\,6.8       & 12.62\,$\pm$\,0.03 & 13.81\,$\pm$\,0.03 & 13.78\,$\pm$\,0.03 & 12.33\,$\pm$\,0.02 & 12.34\,$\pm$\,0.02 & 12.52\,$\pm$\,0.02  \\
2017-11-05   & 63.3     &  +\,8.8       & 12.60\,$\pm$\,0.03 & 13.80\,$\pm$\,0.03 & 13.72\,$\pm$\,0.04 & 12.37\,$\pm$\,0.03 & 12.37\,$\pm$\,0.03 & 12.59\,$\pm$\,0.02  \\
2017-11-10   & 68.0     &  +\,13.5      & 12.61\,$\pm$\,0.03 & 13.81\,$\pm$\,0.03 & 13.72\,$\pm$\,0.04 & 12.48\,$\pm$\,0.03 & 12.44\,$\pm$\,0.03 & 12.73\,$\pm$\,0.02  \\
2017-11-16   & 74.0     &  +\,19.5      & 12.69\,$\pm$\,0.03 & 13.77\,$\pm$\,0.03 & 13.63\,$\pm$\,0.03 & 12.72\,$\pm$\,0.03 & 12.60\,$\pm$\,0.02 & 12.92\,$\pm$\,0.02  \\
2017-11-19   & 77.1     &  +\,22.6      & 12.72\,$\pm$\,0.03 & 13.73\,$\pm$\,0.03 & 13.58\,$\pm$\,0.03 & 12.84\,$\pm$\,0.03 & 12.71\,$\pm$\,0.02 & 13.07\,$\pm$\,0.02  \\
2017-11-22   & 80.1     &  +\,25.6      & 12.77\,$\pm$\,0.04 & 13.77\,$\pm$\,0.03 & 13.58\,$\pm$\,0.04 & 12.96\,$\pm$\,0.03 & 12.81\,$\pm$\,0.03 & 13.23\,$\pm$\,0.03  \\
2017-11-30   & 87.8     &  +\,33.3      & 12.92\,$\pm$\,0.04 & 13.86\,$\pm$\,0.04 & 13.61\,$\pm$\,0.05 & 13.42\,$\pm$\,0.03 & 13.19\,$\pm$\,0.03 & 13.69\,$\pm$\,0.03  \\
2017-12-04   & 92.1     &  +\,37.6      & 13.13\,$\pm$\,0.03 & 13.94\,$\pm$\,0.03 & 13.70\,$\pm$\,0.03 & 13.71\,$\pm$\,0.03 & 13.42\,$\pm$\,0.03 & 13.95\,$\pm$\,0.03  \\
2017-12-08   & 95.7     &  +\,41.2      & 13.22\,$\pm$\,0.04 & 13.98\,$\pm$\,0.03 & 13.68\,$\pm$\,0.05 & --                 & 13.67\,$\pm$\,0.04 & 14.25\,$\pm$\,0.04  \\
2017-12-12   & 99.8     &  +\,45.3      & 13.41\,$\pm$\,0.03 & 14.07\,$\pm$\,0.03 & 13.73\,$\pm$\,0.04 & 14.24\,$\pm$\,0.03 & 13.84\,$\pm$\,0.03 & 14.51\,$\pm$\,0.03  \\
\hline
\end{tabular} \\
$^{a}$ With reference to the maximum light JD~2458054.5. \\
\end{table*}

\begin{table*}
\centering
\caption{Details of 17 isolated field stars within 5$\degr$ radius around the SN which are used to constrain
the interstellar polarization (Section~\ref{isp_cor}). Listed distances are taken from \textit{Gaia} DR2 
catalogue. The mentioned polarization values for HD and S$_{1-4}$ star IDs are from Heiles catalogue and 
estimated from AIMPOL observations, respectively.
\label{tab:field_stars}}
\begin{tabular}{lccccccc}
\hline \hline
Star   &  RA (J2000) & Dec (J2000) & Distance & $P_{V} \pm \sigma_{P_{V}}$ & $ \theta{_V} \pm \sigma_{\theta{_V}}$ & $Q$          & $U$      \\
ID     & (h:m:s)     & (d:m:s)     & (pc)     & (per cent)                 & ($^\circ$)                            &  (per cent ) & (per cent) \\
\hline
HD~693         & 00:11:16.1 & --\,15:27:51.1 &  17.99 $\pm$ 0.11  & 0.01 $\pm$ 0.01 & 135 $\pm$ 29 & --\,0.0000 $\pm$ 0.0082 & --\,0.0080 $\pm$ 0.0090 \\
HD~1388        & 00:17:58.6 & --\,13:27:20.2 &  26.94 $\pm$ 0.04  & 0.02 $\pm$ 0.02 & 136 $\pm$ 25 &     0.0003 $\pm$ 0.0168 & --\,0.0190 $\pm$ 0.0180 \\
HD~967         & 00:14:04.2 & --\,11:18:40.0 &  42.84 $\pm$ 0.10  & 0.11 $\pm$ 0.05 & 108 $\pm$ 13 & --\,0.0868 $\pm$ 0.0507 & --\,0.0626 $\pm$ 0.0504 \\
HD~224383      & 23:57:33.3 & --\,09:38:49.6 &  51.37 $\pm$ 0.17  & 0.02 $\pm$ 0.03 & 115 $\pm$ 41 & --\,0.0102 $\pm$ 0.0252 & --\,0.0123 $\pm$ 0.0261 \\
HD~1195        & 00:16:18.6 & --\,14:28:18.8 &  79.11 $\pm$ 0.44  & 0.02 $\pm$ 0.01 & 134 $\pm$ 12 & --\,0.0007 $\pm$ 0.0089 & --\,0.0210 $\pm$ 0.0090 \\
HD~1064        & 00:14:54.5 & --\,09:34:10.2 & 110.49 $\pm$ 1.36  & 0.07 $\pm$ 0.01 & 141 $\pm$  5 &     0.0136 $\pm$ 0.0111 & --\,0.0676 $\pm$ 0.0110 \\
HD~223774      & 23:52:30.1 & --\,14:15:04.3 & 112.70 $\pm$ 1.97  & 0.11 $\pm$ 0.06 & 150 $\pm$ 14 &     0.0553 $\pm$ 0.0570 & --\,0.0974 $\pm$ 0.0577 \\
HD~604         & 00:10:26.1 & --\,14:27:05.0 & 132.17 $\pm$ 0.89  & 0.09 $\pm$ 0.01 & 135 $\pm$  4 &     0.0010 $\pm$ 0.0140 & --\,0.0910 $\pm$ 0.0140 \\
HD~1231        & 00 16 34.1 & --\,13:24:12.2 & 150.19 $\pm$ 1.39  & 0.11 $\pm$ 0.01 &  17 $\pm$  3 &     0.0924 $\pm$ 0.0120 &     0.0633 $\pm$ 0.0121 \\
HD~224481      & 23:58:21.2 & --\,15:50:50.6 & 164.00 $\pm$ 1.91  & 0.10 $\pm$ 0.04 & 121 $\pm$ 10 & --\,0.0469 $\pm$ 0.0347 & --\,0.0883 $\pm$ 0.0349 \\
HD~223559      & 23:50:33.3 & --\,14:24:05.0 & 181.42 $\pm$ 6.67  & 0.17 $\pm$ 0.03 & 147 $\pm$  6 &     0.0667 $\pm$ 0.0340 & --\,0.1542 $\pm$ 0.0340 \\
HD~1114        & 00:15:24.4 & --\,13:03:15.1 & 210.34 $\pm$ 2.54  & 0.16 $\pm$ 0.02 & 136 $\pm$  3 &     0.0079 $\pm$ 0.0170 & --\,0.1618 $\pm$ 0.0170 \\
HD~1268        & 00:16:52.9 & --\,14:43:03.7 & 259.07 $\pm$ 5.85  & 0.16 $\pm$ 0.02 & 145 $\pm$  3 &     0.0529 $\pm$ 0.0169 & --\,0.1520 $\pm$ 0.0170 \\
\hline
S$_1$$^{\dagger}$ & 00:03:47.5 & --\,11:28:35.1 & 3640.3 $\pm$ 551.3 & 0.52 $\pm$ 0.02 & 159 $\pm$  1 &  0.3822 $\pm$ 0.0184   & --\,0.3482 $\pm$ 0.0181 \\
S$_2$$^{\dagger}$ & 00:03:51.2 & --\,11:27:04.4 & 1588.1 $\pm$  86.2 & 0.29 $\pm$ 0.21 &  61 $\pm$ 15 &--\,0.1519 $\pm$ 0.1728 &  0.2517    $\pm$ 0.1980 \\
S$_3$$^{\dagger}$ & 00:04:00.9 & --\,11:26:47.7 &  549.1 $\pm$   9.9 & 0.24 $\pm$ 0.22 &   5 $\pm$ 18 &  0.2364 $\pm$ 0.2144   &  0.0415    $\pm$ 0.1544 \\
S$_4$$^{\dagger}$ & 00:04:04.3 & --\,11:30:02.6 & 1843.0 $\pm$ 131.1 & 0.65 $\pm$ 0.39 & 165 $\pm$ 15 &  0.5647 $\pm$ 0.3773   & --\,0.3260 $\pm$ 0.3483 \\
\hline
\end{tabular} \\
$^{\dagger}$ Field stars around SN~2017hcc within the FOV of AIMPOL.\\
\end{table*}

Imaging polarimetric observations of SN 2017hcc were also made using the Hiroshima Optical and Near-InfraRed 
camera \citep[HONIR;][]{2014SPIE.9147E..4OA} mounted on the 1.5-m Kanata telescope at Higashi-Hiroshima Observatory,  
in the $R$-band ($\lambda_{R_{eff}}$ = 0.65 $\mu$m) on 2017 November 9.5 and December 22.5. The observations 
consisted of a sequence of exposures at four position angles of the achromatic HWP (i.e. $0^{\circ}.0, 45^{\circ}.0, 
22^{\circ}.5$, and $67^{\circ}.5$).
The position angle of the polarization was calibrated using the observations of the polarized standard stars
\citep[BD$+$59$^{\circ}$389, BD$+$64$^{\circ}$106;][]{1992AJ....104.1563S} as listed in Table~\ref{stand_log}. 
Since the instrumental polarization was estimated to be negligible ($\sim$\,0.02 per cent), it has not been 
corrected for.

The Ultra Violet Optical Telescope (UVOT) on-board the \textit{Swift} satellite monitored SN~2017hcc from 
JD~2458054.9 to 2458099.8 in $uvw2$, $uvm2$, $uvw1$, $u$, $b$ and $v$ bands \citep[see][for filter
specifications]{2008MNRAS.383..627P}. These data sets were retrieved from the \textit{Swift} data archive.  
To estimate the SN magnitude in each band, aperture photometry was performed using various modules of 
the {\small HEASOFT} (High Energy Astrophysics Software). A detailed description can be found in 
\citet{2018MNRAS.473.3776K} and \citet{2018MNRAS.475.2591S}. The UVOT magnitudes (in Vega-system) are 
listed in Table~\ref{swift_mag}.

We obtained a low-resolution spectrum of SN~2017hcc with the Himalayan Faint Object Spectrograph Camera (HFOSC), 
mounted on the f/9 Cassegrain focus of the 2-m Himalayan Chandra Telescope (HCT) of Indian Astronomical Observatory 
(IAO), Hanle, India. Observations were performed on 2017 November 23 ($\sim$\,27 d post-maximum) using two grisms 
Gr\#7 (3500\,--\,7800 \AA) and Gr\#8 (5200\,--\,9250 \AA), having a resolution of $\sim$\,7 \AA. A description on 
the data reduction procedure can be found in our previous studies \citep[e.g.][]{2018MNRAS.473.3776K,2018MNRAS.475.2591S,
2018MNRAS.480.2475S,Avinash_sn16gfy}. The spectrum was brought to an absolute flux scale using the blackbody fit to the 
extinction ($E(B-V)$ = 0.029 mag) corrected \textit{Swift} UVOT fluxes.

\begin{figure*}
\centering
\includegraphics[scale = 0.18]{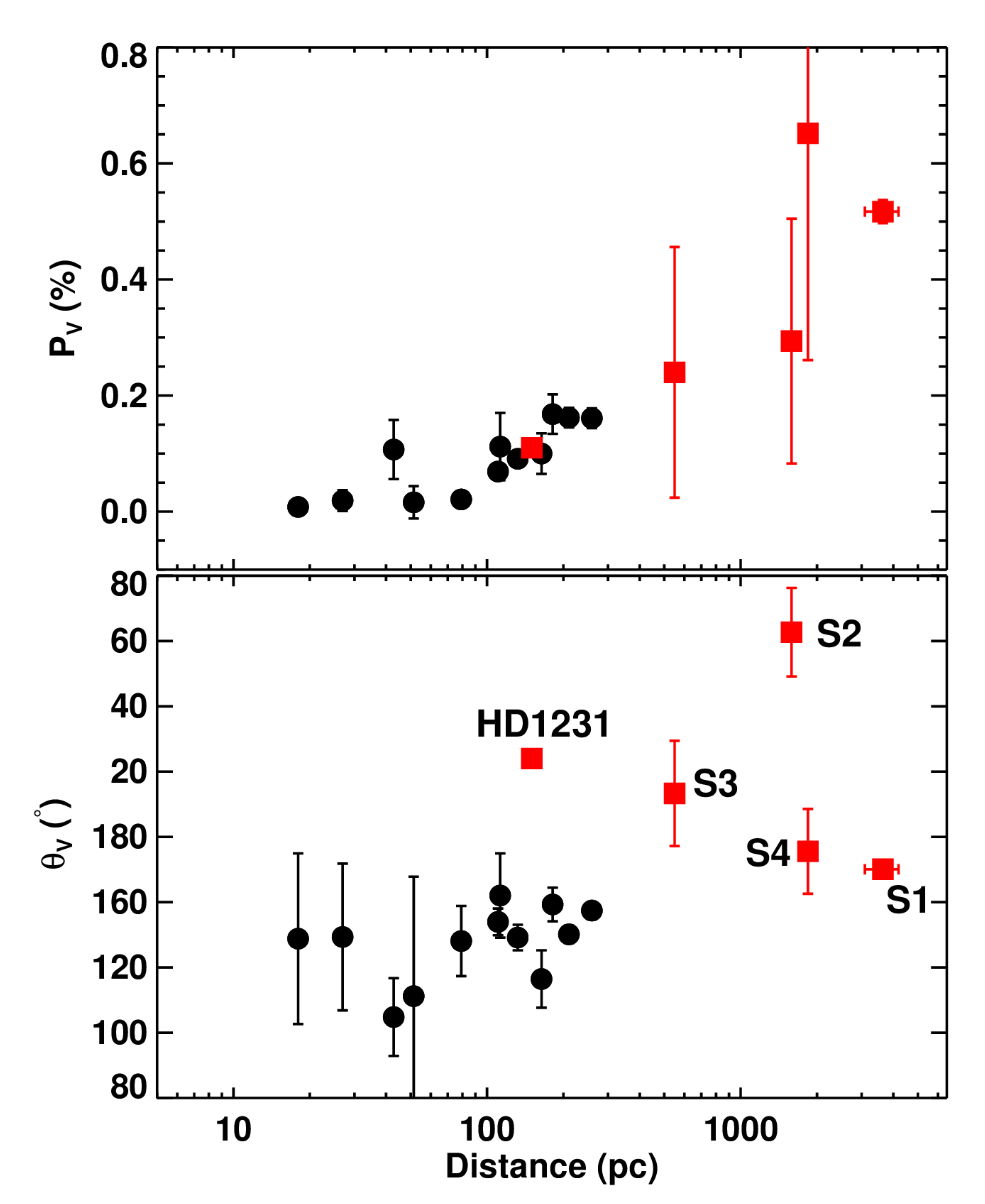}
\includegraphics[scale = 0.18]{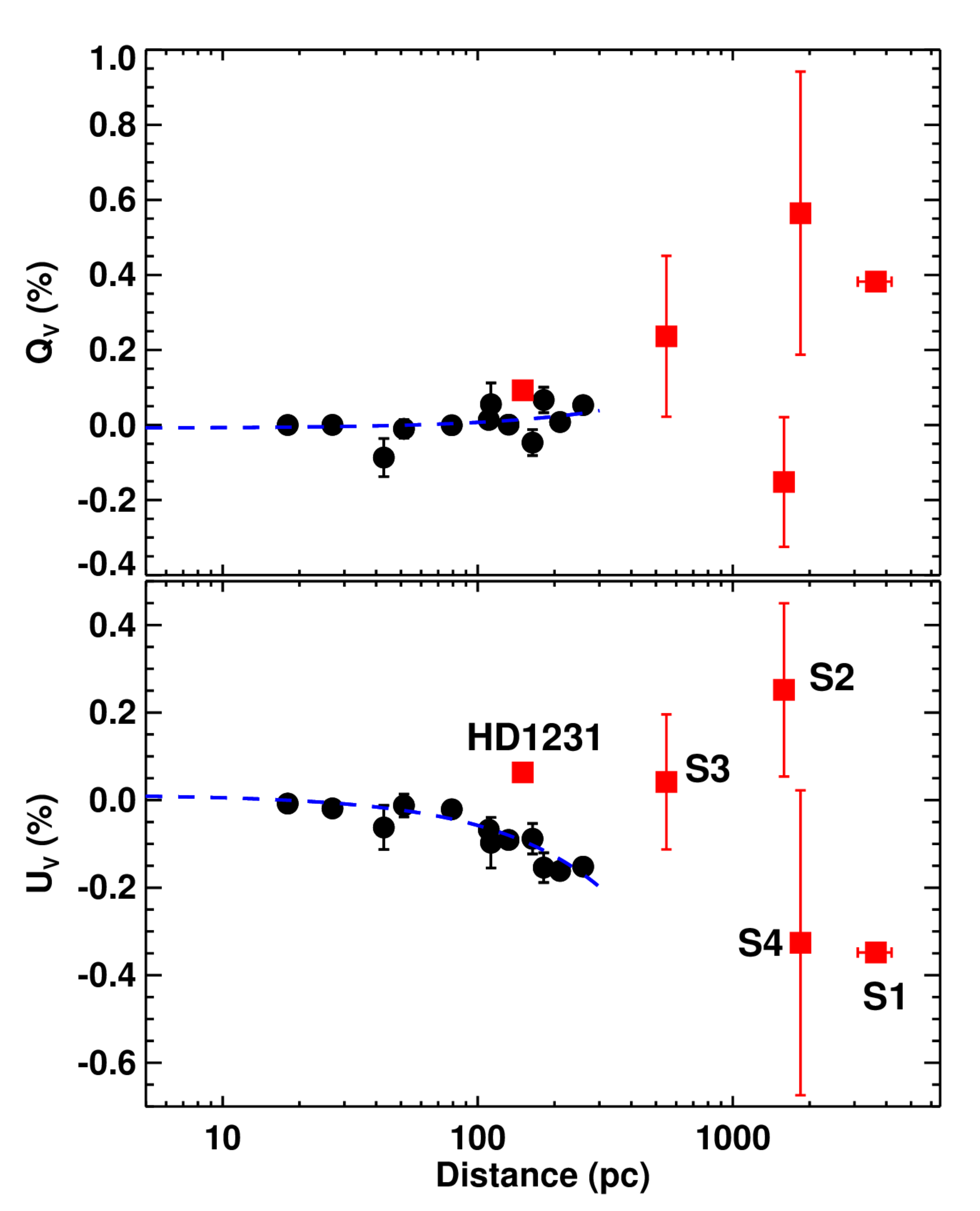}
\caption{Distribution of polarization parameters of 17 field stars around SN~2017hcc.
Left panel: $P_{V}$ and $\theta_{V}$ are plotted against distance in top and bottom panels, 
respectively. The polarization values are from \citet{2000AJ....119..923H} (dot symbols 
and HD~1231) and AIMPOL (S$_{1-4}$). Parallax measurements are from \citet{2018A&A...616A...1G}.
Right panel: $Q_{V}$ and $U_{V}$ plots for same field stars. In each panel, dashed lines 
are the weighted linear fits on the Stokes parameters vs distances. Notably, HD~1231 and 
S$_{1-4}$ (indicated with filled squares) were not used in the fits (see Section~\ref{isp_cor}).}
\label{heiles_aimpol}
\end{figure*}

\section{Results and discussion}\label{result}

In this section, we constrain the interstellar polarization (ISP) and derive the intrinsic polarization 
characteristics of SN~2017hcc. The evolution of light curves, spectral properties and physical parameters
are also discussed.

\subsection{Interstellar polarization}\label{isp_cor}

The interstellar polarization originates (both due to the Milky Way (MW) and/or the host galaxy) 
owing to directional extinction of the dust grains situated along the line-of-sight of an observer. 
Accordingly, the intrinsic signal of a source gets altered when it passes through 
the interstellar medium. The situation becomes more complicated in case of extragalactic SNe as 
the host galaxy dust and local environment may further contaminate the propagating signal. 
Consequently, the intervening ISP contribution should be accounted for from the observed SN 
measurements in order to derive the intrinsic polarization characteristics of SN.
While temporal variation may be observed in the SN polarization properties, the ISP should be 
constant towards a particular direction. Therefore, if ISP can be constrained appropriately 
(or at-least setting an upper limit), it will be easier to estimate intrinsic SN parameters by 
subtracting it from the observed values. Fortunately, the Galactic ISP has been properly documented 
in literature \citep*[e.g.][]{1975ApJ...196..261S,1978A&A....66...57W}.

In order to examine the ISP towards the SN arising due to the MW, the nearby field stars (well above the Galactic 
plane) can serve as good probe \citep{1995ApJ...440..565T}. For this purpose, we selected 13 isolated field 
stars within a radius of 5$^\circ$ around the location of SN~2017hcc which do not show emission, variability 
and double characteristics in SIMBAD database. The polarization and distance information of these sources 
were collected from the polarization data catalog of \citet{2000AJ....119..923H} and the newly 
available \textit{Gaia} data (DR2) \citep{2018A&A...616A...1G}, respectively. Additionally, 4 stars 
(indicated with IDs 1\,--\,4 in Fig.~\ref{fig_1}) lying within the 8 arc min AIMPOL field of SN~2017hcc were 
also considered. Different parameters of all 17 stars are listed in Table~\ref{tab:field_stars} and their 
$P_{V}$ and $\theta_{V}$ measurements are plotted against distance in Fig.~\ref{heiles_aimpol} (left panel).  
An increasing trend can be seen in the $P_{V}$ values as a function of distance up to 
$\sim$\,260 pc. The $\theta_{V}$ values are generally distributed between 100$^\circ$ and 150$^\circ$ (except 
for a few sources).

The $P_{V}$ and $\theta_{V}$ of the field stars were converted into Stokes parameters ($Q$ and $U$) using 
the relations $Q_{V}$ = $P_{V}$\,cos(2$\theta_{V}$) and $U_{V}$ = $P_{V}$\,sin(2$\theta_{V}$). The distribution 
of $Q_{V}$ and $U_{V}$ with respect to distance is shown in Fig.~\ref{heiles_aimpol} (right panel). 
Further, weighted linear fits on the distance versus Stokes parameters were performed as shown with dotted 
lines in the same figure. 
It is to be noted that the Heiles catalogue stars have reliable information on their spectral types. 
However, this is not the case for field stars situated in the field-of-view (FOV) of AIMPOL 
\citep[except for the RR Lyrae variable, S1,][]{2014MNRAS.441..715G}. Stars S2, S3 and S4 are faint and have 
large polarization error bars due to low SNR. HD~1231 is deviant amongst the Heiles catalogue stars. 
Consequently, these stars (S1-S4 and HD~1231) were excluded from the fitting.

The fitted slopes and intercepts at the distance 260 pc of HD 1268 were used to estimate the Stokes 
parameters $Q_{ISP_{MW}}$ and $U_{ISP_{MW}}$ values corresponding to the ISP contribution in the MW. 
From these parameters, the derived $P_{ISP_{MW}}$ and $\theta_{ISP_{MS}}$ were found to be 0.17 $\pm$ 
0.02 per cent and 140.3$^\circ$ $\pm$ 3.0$^\circ$, respectively. 

The MW reddening in the direction of \hcc\ was estimated as $E(B-V)$ = 0.029 $\pm$ 0.001 mag using the 
all-sky dust extinction map \citep*{2011ApJ...737..103S}.
Polarimetric studies of Galactic stars along different directions have provided a correlation between ISP 
and reddening \citep[e.g.][]{1975ApJ...196..261S}. According to this, the mean and maximum polarization 
efficiency are correlated with interstellar extinction as $P_{mean} = 5 \times E(B-V)$ and $P_{max} = 
9 \times E(B-V)$, respectively. Assuming that MW dust grains follow mean polarization efficiency along 
the SN direction, the first relation yields $P_{mean}$ of 0.15 per cent. This value is similar to the 
ISP value ($P_{ISP}$ = 0.17 per cent) computed using the polarization measured for the field stars. 
Using the maximum polarization efficiency relation, an upper limit of i.e. $P_{max}$ = 0.26 per cent 
is obtained.

The 3D dust extinction map\footnote{\url{http://argonaut.skymaps.info/}} \citep{2018MNRAS.478..651G} was used 
to further explore the MW dust extinction along the line-of-sight towards SN~2017hcc. The map is based on 
high-quality stellar photometry data from Pan-STARRS 1 and 2MASS surveys, and provides information of $E(B-V)$ 
as a function of distance. The extinction map indicates the best fit reddening value remains constant around 
0.033 mag beyond a distance of 650 pc. This implies the predicted degree of polarization, as per the mean 
polarization efficiency relation, would be 0.17 per cent. This again confirms our ISP estimate using the 
weighted linear fits on distance versus Stokes parameters. 

The colour excess of the host-galaxy can be inferred using the equivalent width (EW) of Na\,{\sc i} D absorption
lines \citep{2003fthp.conf..200T,2012MNRAS.426.1465P,2018A&A...609A.135S}. This signature is very weak at the 
redshift of the host galaxy in the low-resolution spectrum presented here (see inset in Fig.~\ref{spec_comp}). 
We estimate an upper limit of 0.06 \AA\ on Na\,{\sc i} D EW, indicating negligible host extinction. 
\citet{2017ATel10911....1M} have made a similar observation based on spectropolarimetric data obtained close 
to maximum light. 

The direct and indirect methods discussed above, indicate that the MW aligned dust grains exhibit mean polarization 
efficiency. Therefore, in the present analysis we have considered only MW dust as the major source of ISP along the 
LOS of SN~2017hcc. An average ISP as a function of wavelength was estimated using the formulation of 
\citet{1975ApJ...196..261S} and found to be similar for both $V$ and $R$ bands ($P_{ISP}$ = 0.17 per cent and 
$\theta_{ISP}$ = 140.3$^\circ$). The corresponding $Q_{ISP}$ and $U_{ISP}$ values are 0.032 $\pm$ 0.018 and --\,0.170 
$\pm$ 0.019 per cent which are subtracted from the observed SN polarization measurements (see Section~\ref{int_pol}). 
A similar methodology has been used to derive the ISP contribution in dark globules and B[e] stars 
\citep{2018AJ....156..115L, 2019ApJ...875...64E}.

\begin{figure}
\centering
\includegraphics[width=\columnwidth]{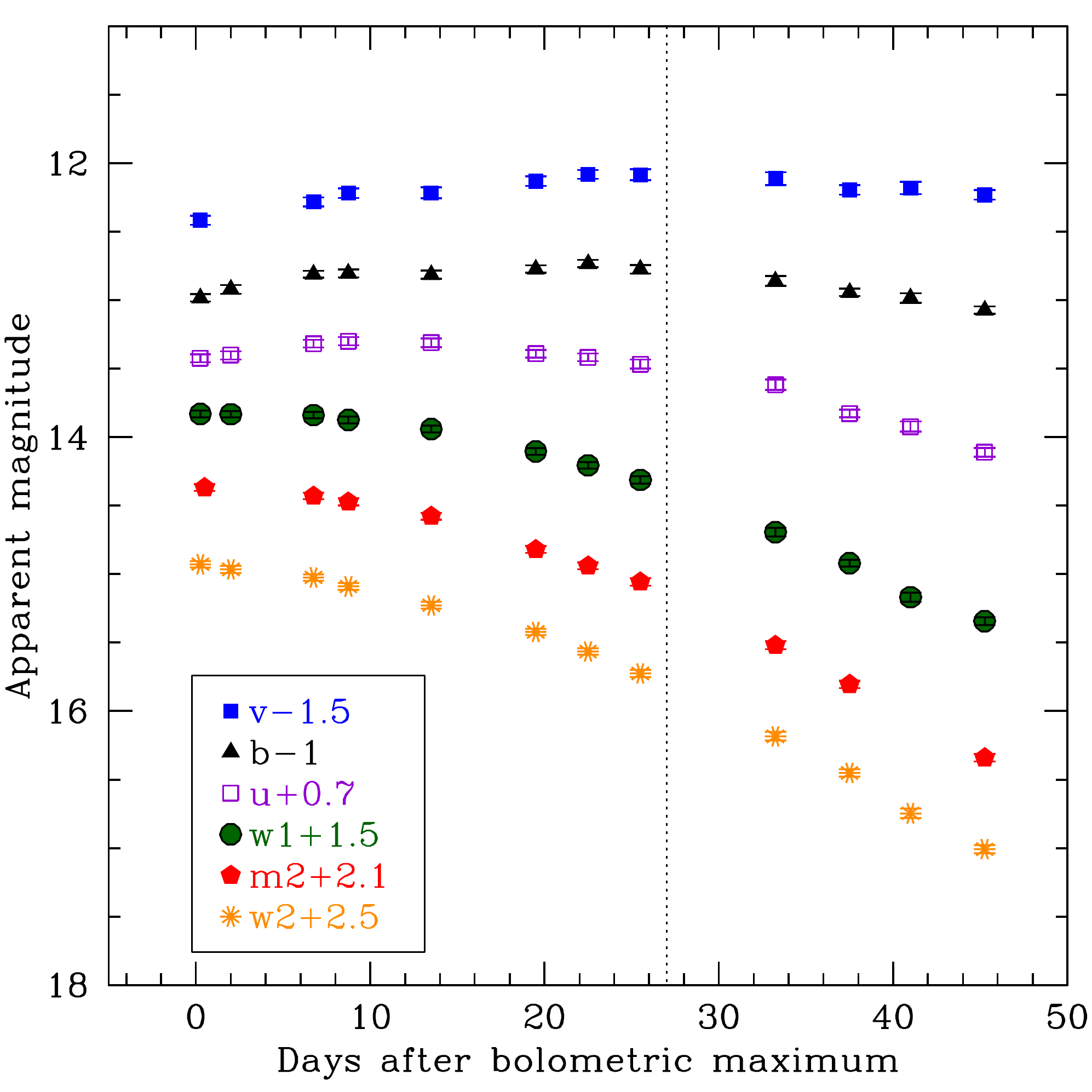}
\caption{Evolution of \textit{Swift}-UVOT light curves. Vertical shift with indicated amount has been applied for clarity.
A dotted line indicates epoch of spectrum taken with the HCT (see Section~\ref{obs} and \ref{comp}).}
\label{swift_lc}
\end{figure}

\begin{figure*}
\centering 
\includegraphics[scale=0.5]{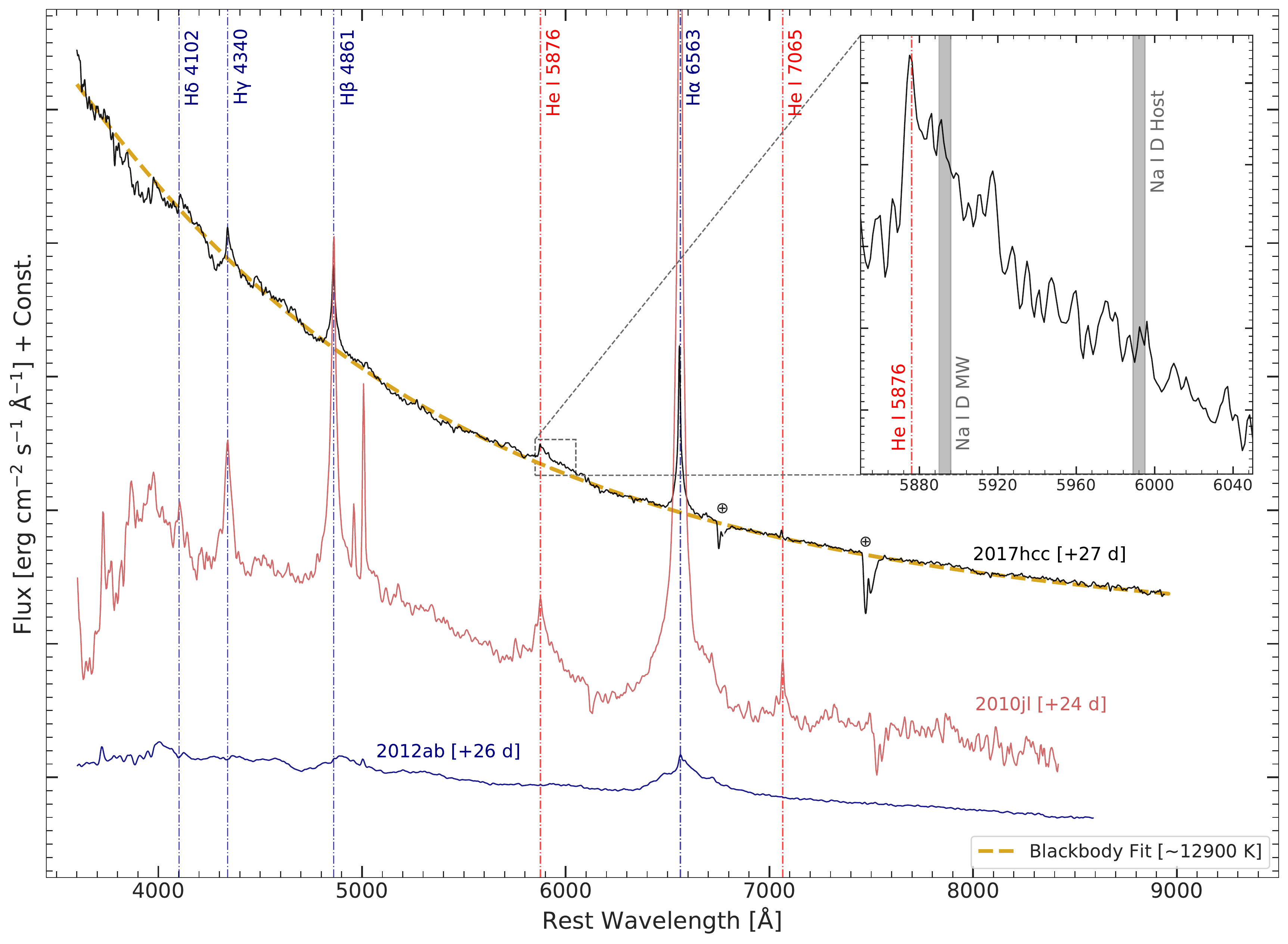}
\caption{Comparison of SN~2017hcc, SN~2012ab \citep{Anjasha} and SN~2010jl \citep{2012AJ....144..131Z} spectra at
similar epochs. Prominent Balmer and He\,{\sc i} lines are also marked. The encircled plus symbols indicate telluric
lines. The spectra have been corrected for the respective host galaxy redshift. Given phases are relative to maximum
light. The blackbody model fit to SN~2017hcc is shown with dotted line (see Section~\ref{comp}). A zoomed spectrum 
region (5850\,--\,6050 \AA) of SN~2017hcc is shown in the inset.}
\label{spec_comp}
\end{figure*}

\subsection{Light curves and spectral features}\label{comp}

The \textit{Swift} UVOT light curves of SN~2017hcc are displayed in Fig.~\ref{swift_lc}. We adopt JD~2458054.5 as the 
epoch of maximum from \citet{2017RNAAS...1a..28P} as inferred from their bolometric light curve. The $b$ and $v$ band 
light curves are almost flat compared to the bluer bands. The decay rates (between 0 to 45 d) in $u$, $uvw1$, $uvw2$ 
and $uvm2$ bands are $\sim$\,0.02, 0.04, 0.05 and 0.05 mag d$^{-1}$, respectively. 

\begin{figure}
\centering
\includegraphics[width=\columnwidth]{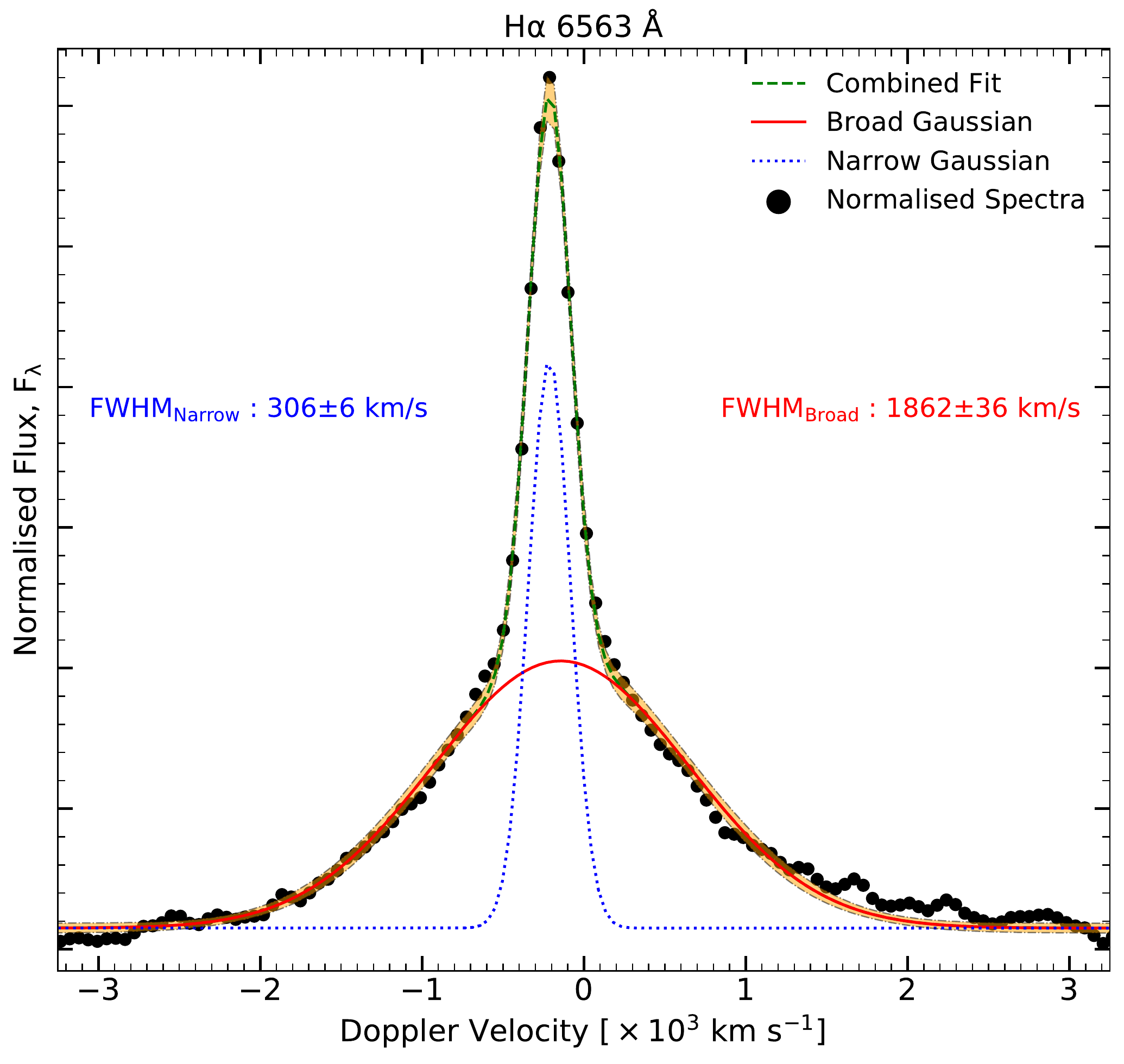}
\caption{Gaussian fit to the observed spectrum of SN~2017hcc.
The 3$\sigma$ confidence interval of the fit has been indicated with a orange shaded region.}
\label{fig_gaus_fit}
\end{figure}

In Fig.~\ref{spec_comp}, the redshift corrected \citep[$z$ = 0.0168,][]{2017RNAAS...1a..28P} spectrum of SN~2017hcc, 
obtained on 2017 November 23 ($\sim$\,+27 d) is displayed. Spectra of SN~2010jl \citep{2012AJ....144..131Z} and 
SN~2012ab \citep{Anjasha}, at a similar epoch are also shown in the Figure for comparison. It is evident from the 
figure that the spectrum of SN~2017hcc is considerably bluer in comparison to the other two events. 
\citet{2017RNAAS...1a..28P} obtained a spectrum of SN~2017hcc on 2017 October 20 ($\sim$\,+8 d) which also reported
a significantly bluer continuum.
The primary spectral features are also marked in Fig.~\ref{spec_comp}. The emission lines of hydrogen
i.e. H$\alpha$ (6563 \AA), H$\beta$ (4861 \AA), H$\gamma$ (4341 \AA) and H$\delta$ (4102 \AA) are
prominent in all the three events but the H$\alpha$ and H$\beta$ lines are broader in SN~2012ab and
SN~2010jl compared to SN~2017hcc. The blackbody fit to the continuum indicates a temperature of 
$\sim$\,12900 $\pm$ 680 K (shown with a dotted line in Fig.~\ref{spec_comp}).
The flux ratio of H$\alpha$ to H$\beta$ lines is estimated as $\sim$\,4 for SN~2017hcc whereas the 
same ratio is found to be $\sim$\,5 and $\sim$\,8 for SN~2010jl and SN~2012ab, respectively. 
The H$\alpha$ line profile is decomposed into narrow and broad components by fitting a two-component 
Gaussian as shown in Fig.~\ref{fig_gaus_fit}. The best fit yields an FWHM of 306 km s$^{-1}$ and 1862 
km s$^{-1}$ for the narrow and broad components, respectively.

The He\,{\sc i} 5876, 7065\, \AA\ features are stronger in SN~2017hcc and SN~2010jl than in SN~2012ab.
An unblended signature of a Na\,{\sc i} D feature was found in SN~2012ab \citep{2018MNRAS.475.1104B}
which is negligible in SN~2017hcc and SN~2010jl \citep{2012AJ....144..131Z} indicating
negligible reddening (see Section~\ref{isp_cor}) along the line of sight caused by the host galaxies.

\subsection{Basic physical parameters}\label{prog}

\begin{figure}
\centering
\includegraphics[width=\columnwidth]{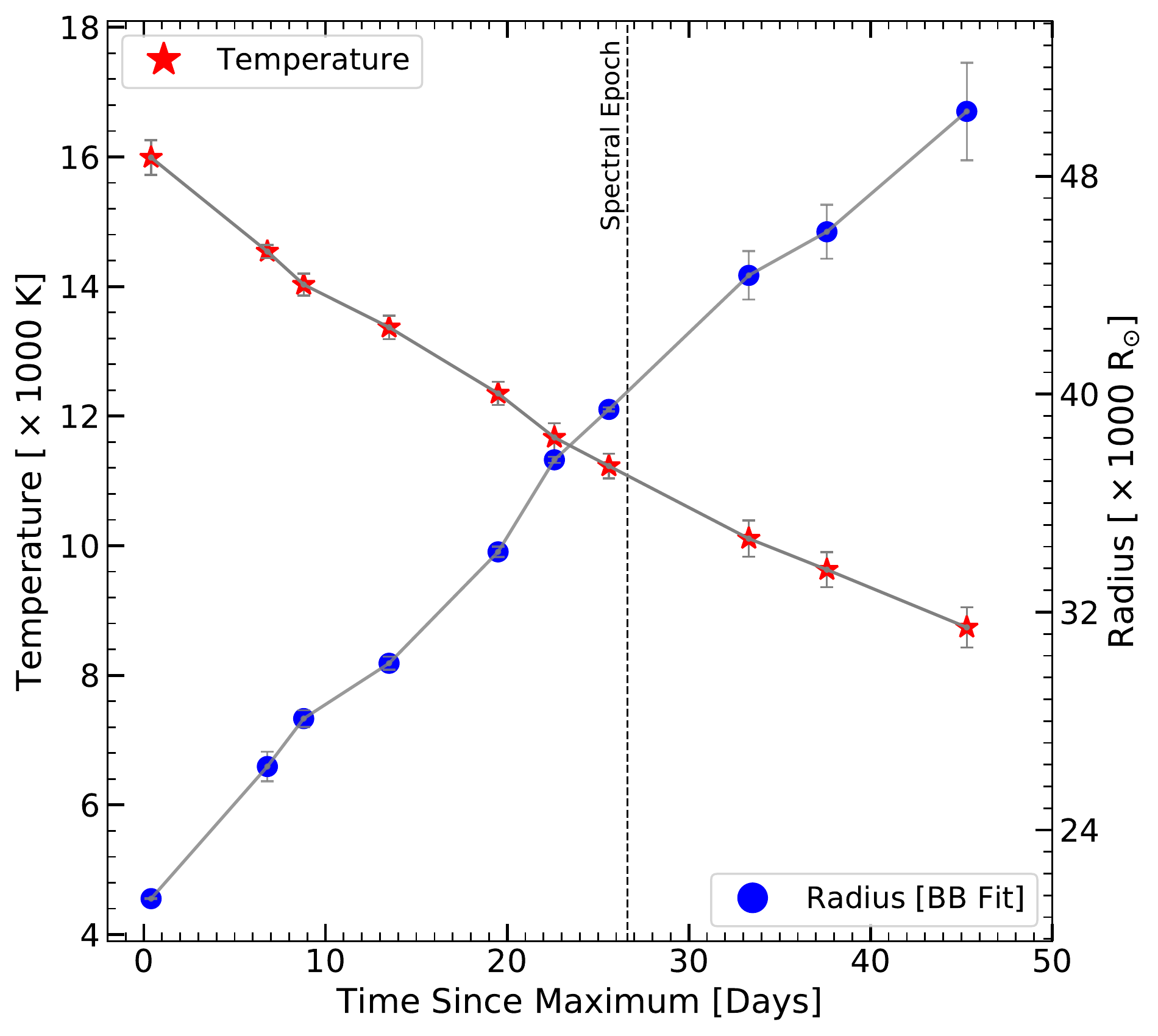}
\caption{Temperature and radius evolution of SN~2017hcc.}
\label{BB}
\end{figure}

The temperature and radius evolution of SN~2017hcc was computed by fitting a blackbody function to the Spectral 
Energy Distribution (SED) constructed using the {\it Swift} magnitudes (c.f. Table~\ref{swift_mag}). Weighted 
Least-Squares optimization was employed for the fit with the weights being the inverse of the 1$\sigma$ error on the 
{\it Swift} magnitudes. The evolution of these parameters are shown in Fig.~\ref{BB}. The temperature at the earliest 
epoch ($\sim$\,0 d), estimated at $\sim$\,16000 K is similar to the temperature obtained in \citet{2017RNAAS...1a..28P}. 
The temperature subsequently decreased to $\sim$\,8700 K at the last epoch ($\sim$\,+45 d). In comparison, SN~2012ab had 
a temperature of $\sim$\,12000 K at $-22$ d, which decreased to $\sim$\,10500 K close to the maximum. The peculiar Type 
IIn SN~2006gy displayed a temperature of 12000 K \citep{2007ApJ...666.1116S} whereas SN~2006tf reached a temperature 
of $\sim$\,7800 K near the peak \citep{2008ApJ...686..467S}.

The temperature evolution of SN~2017hcc clearly indicates that it is hotter than other Type IIn events at similar 
epochs. The blackbody radius of Type IIn SNe is found to steadily increase with time, reach a peak and subsequently 
decline, as observed in several events like SN~1998S \citep{2001MNRAS.325..907F}, SN~2005gj \citep{2007arXiv0706.4088P} 
and SN~2006gy \citep{2010ApJ...709..856S}. The peak represents a transition to the optically thin CSM shell 
\citep{2010ApJ...709..856S}. In SN~2017hcc, the blackbody radius was found to be increasing until the last epoch of 
{\it Swift} observation (c.f. Fig.~\ref{BB}) which is likely due to the optically thick SN atmosphere until day +45. 

The overall light curve evolution and spectral features suggests the presence of CSM interaction in SN~2017hcc. 
Assuming that the luminosity is mainly powered by the ejecta-CSM interaction \citep{1994ApJ...420..268C}, the 
mass-loss rate ($\dot M$) can be estimated using the relation of \citet{1994MNRAS.268..173C}: 
\begin{equation}\label{eq_m}
\dot{M}=\frac{2L}{\epsilon}\frac{v_w}{v_{s}^{3}}
\end{equation}
where $L$ is the observed luminosity at a particular epoch, $\epsilon$ represents the efficiency of converting 
the shock's kinetic energy into visual light. $v_w$ and $v_s$ are the velocities of the pre-explosion stellar 
wind and the post-shock shell, respectively. We adopt a magnitude of $M_v = -20.8 \pm 0.1$ on day +27 by 
interpolating the \textit{Swift} magnitude provided in Table~\ref{swift_mag} and estimate the luminosity to be 
$L$ = $\sim$\,6.0 $\times$ 10$^{42}$ erg s$^{-1}$. No bolometric correction has been applied in this calculation. 
Assuming a constant value of $v_w = 20$ km s$^{-1}$ for a Red Supergiant (RSG) wind and $v_s = 1862$ km s$^{-1}$ 
for the SN ejecta (see Section~\ref{comp}). For a value of $\epsilon$ = 0.5 \citep[though, $\epsilon$ is optical 
depth dependent,][]{2007ApJ...671L..17S}, $\dot M$ is estimated as 0.12 M$_{\sun}$ yr$^{-1}$. The typical mass-loss 
rates in RSGs, yellow hypergiants and luminous blue variables (LBVs) are found to have a range of 
10$^{-6}$\,--\,10$^{-4}$ M$_{\sun}$ yr$^{-1}$, 10$^{-4}$\,--\,10$^{-3}$ M$_{\sun}$ yr$^{-1}$ and 
0.1\,--\,10 M$_{\sun}$ yr$^{-1}$, respectively \citep[see][]{2014ARA&A..52..487S}. The mass loss rate 
estimated for SN~2017hcc is closer to that seen in LBVs, indicating that its progenitor is likely to be an LBV star.

\begin{table*}
\centering
\caption{Observed and intrinsic polarization parameters of SN~2017hcc.
\label{tab_log}}
\begin{tabular}{ccc|cc|cc|ccc}
\hline \hline
\textsc{ut} Date &JD      & Phase$^{a}$ &Band & \multicolumn{2}{c}{Observed}  & \multicolumn{2}{c}{Intrinsic (ISP subtracted)} & $Q$  &  $U$  \\
(2017)           &2458000 &(Days)       &     & $P \pm \sigma_{P}$    & $ \theta \pm \sigma_{\theta}$ & $P_{int} \pm \sigma_{P_{int}}$ & $\theta_{int} \pm \sigma_{\theta_{int}}$ & (per cent) & (per cent)\\
                 &        &             &     & (per cent)                    & ($^\circ)$                            &  (per cent)
&  ($^\circ$)                      &            &           \\
\hline
Oct 30$^*$ &56.5   &+2.1  & $V$  & 4.84 $\pm$ 0.02 & 96.5 $\pm$  0.1 & 4.84 $\pm$ 0.03 & 95.5 $\pm$ 0.2   & --\,4.747 $\pm$ 0.027 & --\,0.922 $\pm$ 0.026 \\
Dec 25     &113.1  &+58.6 & $V$  & 1.40 $\pm$ 0.10 & 104.2 $\pm$ 2.0 & 1.36 $\pm$ 0.10 & 100.8 $\pm$ 2.2  & --\,1.262 $\pm$ 0.105 & --\,0.498 $\pm$ 0.102 \\
\hline
Nov 9.5    & 67.1  & +12.6 & $R$ & 3.22 $\pm$ 0.13 & 96.41 $\pm$ 1.4 & 3.22 $\pm$ 0.13 & 95.0  $\pm$ 1.4  & --\,3.170 $\pm$ 0.133 & --\,0.553 $\pm$ 0.157 \\
Dec 22.5   & 110.0 & +55.5 & $R$ & 1.39 $\pm$ 0.17 & 104.8 $\pm$ 5.0 & 1.35 $\pm$ 0.20 & 101.4 $\pm$ 4.7  & --\,1.240 $\pm$ 0.191 & --\,0.523 $\pm$ 0.228 \\
\hline
\end{tabular} \\
$^*$ Observation of this epoch has been reported by \citet{2017ATel10911....1M}. \\
$^{a}$ With reference to the maximum light JD~2458054.5. \\
\end{table*}

\begin{figure}
\centering
\includegraphics[width=\columnwidth]{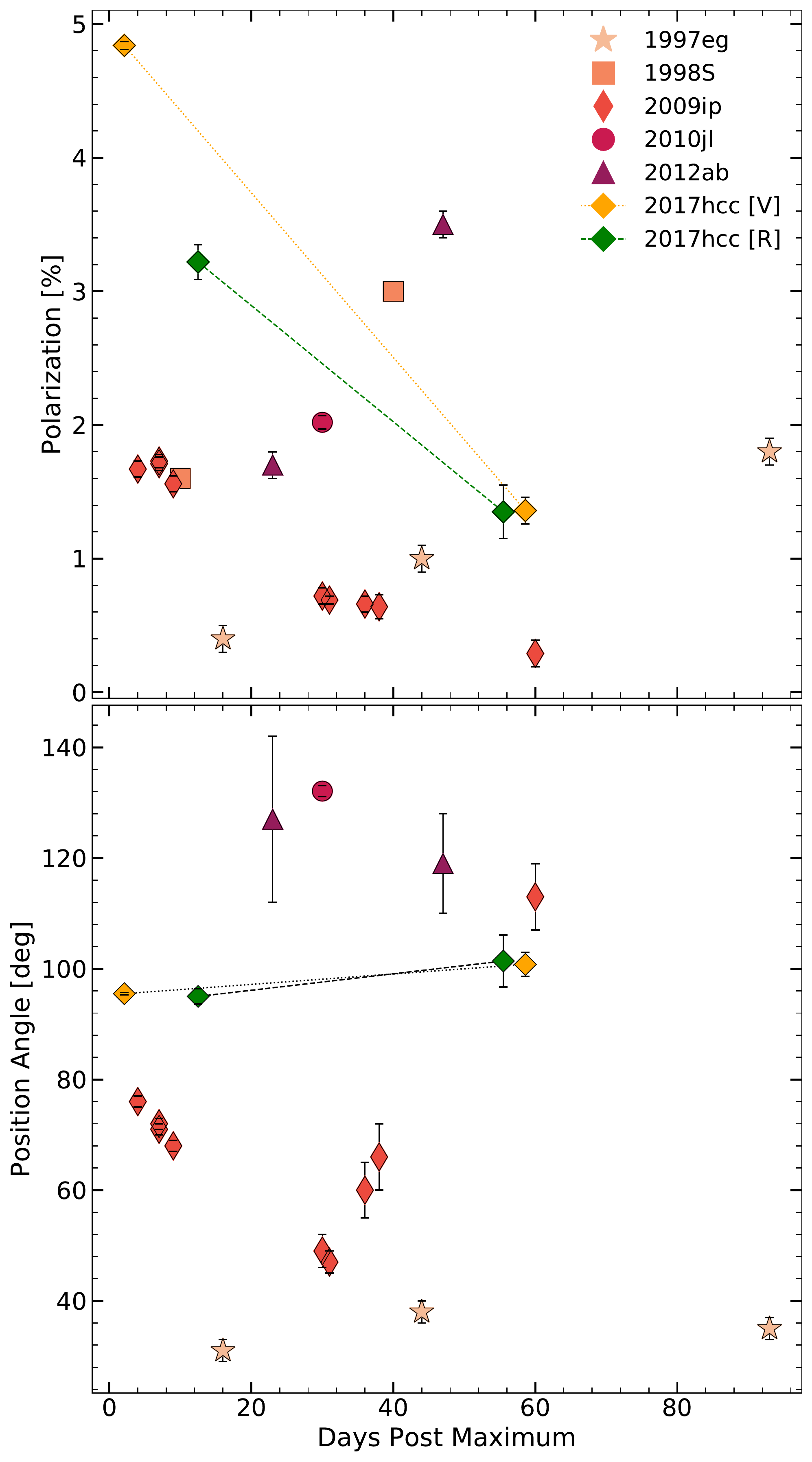}
\caption{Polarization evolution of SN~2017hcc in $V$ and $R$ bands are shown with different symbols and
connected with lines for clarity (Top panel: intrinsic polarization values and bottom panel: position angle). 
Polarization inferred from SN~2012ab \citep{2018MNRAS.475.1104B}, SN~2010jl \citep{2011A&A...527L...6P}, 
SN~2009ip \citep{2014MNRAS.442.1166M}, SN~1998S \citep{2001ApJ...550.1030W} and SN~1997eg 
\citep{2008ApJ...688.1186H} have been plotted for comparison.}
\label{fig_comp}
\end{figure}

\subsection{Intrinsic polarization}\label{int_pol}

To derive the intrinsic polarization parameters of SN~2017hcc, ISP contribution must be subtracted from the 
observed polarization values. The intrinsic Stokes parameters were estimated by the relations: $Q_{int}$ = $Q_{obs}$ 
-- $Q_{ISP}$ and $U_{int}$ = $U_{obs}$ -- $U_{ISP}$. Here $Q_{obs}$, $U_{obs}$ and $Q_{ISP}$, $U_{ISP}$ 
represent the observed and ISP Stokes parameters, respectively. Therefore, at each epoch vectorial subtraction 
was performed and the resulted $Q_{int}$ and $U_{int}$ were converted back to $P_{int}$ and $\theta_{int}$ with 
the help of following equations:
\begin{equation}\label{eq1}
P_{int} = \sqrt{{{Q_{int}}^2} + {{U_{int}}^2}}
\end{equation}
\begin{equation}\label{eq2}
\theta_{int} = 0.5 \times \arctan\left(\frac{U_{int}}{Q_{int}}\right)  
\end{equation}
Both the observed and the intrinsic (ISP corrected) polarization parameters of SN~2017hcc are listed 
in Table~\ref{tab_log} along with the spectropolarimetric observation of \citet{2017ATel10911....1M}. 

The polarization in $V$-band at $\sim$\,+2 d was estimated as 4.84 $\pm$ 0.03 per cent by 
\citet{2017ATel10911....1M}, and a value of 1.36 $\pm$ 0.10 per cent is estimated at $\sim$\,+59 d. 
The $R$-band polarization at $\sim$\,+13 d and $\sim$\,+56 d are estimated at 3.22 $\pm$ 0.13 and 
1.35 $\pm$ 0.20 per cent, respectively. The position angle in both the bands is $\sim$\,100$^\circ$ 
at all epochs. Accounting for the maximum ISP contribution (see Section~\ref{isp_cor}), marginal 
changes are seen in the intrinsic polarization values. Here, a $V$-band polarization of 4.84 per 
cent (95.0$^\circ$) and 1.34 per cent (98.9$^\circ$) were estimated on the $\sim$\,+2 d and $\sim$\,+59 d, 
respectively. Similarly, an $R$-band polarization of 3.22 per cent (94.1$^\circ$) and 1.33 per cent 
(99.5$^\circ$) were estimated on $\sim$\,+13 d and $\sim$\,+56 d, respectively. This clearly indicates that 
the observed polarization values are intrinsic, and the degree of polarization decreased by $\sim$\,3.5 
per cent in $V$-band, and by $\sim$\,2.3 per cent in the $R$-band within a period of $\sim$\,2 months. 
No significant variation is observed in the polarization angle in both the bands during this period.

\subsection{Comparison of SN~2017hcc with other Type IIn events}\label{comp_17hcc}

Only a handful of Type IIn SNe other than \hcc\ have been monitored polarimetrically close to the maximum.
These are SN~1997eg \citep{2008ApJ...688.1186H}, SN~1998S \citep{1998IAUC.6829....1L}, SN~2009ip 
\citep{2014MNRAS.442.1166M,10.1093/mnras/stx1228}, SN~2010jl \citep{2010CBET.2532....1N} and SN~2012ab 
\citep{2012CBET.3022....1V}. The evolution of polarization parameters of SN~2017hcc along with the above 
events is shown in Fig.~\ref{fig_comp}. In general, the temporal evolution in the intrinsic polarization 
of the comparison sample (except SN~2009ip) display an increasing trend with no significant variation in 
their polarization angle. In the case of SN 1997eg, \citet{2008ApJ...688.1186H} discussed the polarization 
properties of SN~1997eg in the context of three ISP values. Their preferred ISP subtracted polarization 
exhibited an increasing trend whereas the other two ISP values resulted in decreasing trend of the intrinsic 
polarization \citep[see Figure~11 in][]{2008ApJ...688.1186H}.

The peculiar SN~2009ip (2012b outburst, \citealp{2013MNRAS.430.1801M,2013ApJ...763L..27P, 2014MNRAS.438.1191S})
displayed a high degree of polarization ($P_V$\,$\sim$\,1.7 per cent) at maximum ($M_V$\,$\sim$\,--18.5 mag) and 
followed a decreasing trend similar to the decay in its light curve, until day $\sim$\,70. This behaviour is 
indicative of a correlation between the intrinsic polarization and photometric luminosity. A similar trend is 
seen in \hcc, with a higher intrinsic polarization ($P_V$\,$\sim$\,4.84 per cent), and a brighter maximum 
($M_V$\,$\sim$\,--20.7 mag). Further, there is almost no variation seen in the polarization angle in the case 
of \hcc, whereas, a substantial variation ($\sim$\,30$^{\circ}$) was seen in the case of SN~2009ip 
\citep{2014MNRAS.442.1166M,10.1093/mnras/stx1228}.

The emergence of polarization among Type IIn SNe is generally attributed to either electron scattering in an
asymmetric photosphere, or asymmetries in the interaction region. In several studies, departure of the SN ejecta 
from spherical symmetry is a favoured explanation. This line of reasoning makes it rather straightforward to explain 
the temporal increase of intrinsic polarization as the core becomes progressively exposed. This also suggests that 
the inner layers of the ejecta are quite aspherical, indicating an aspherical explosion \citep[see][and references 
therein]{1996ApJ...467..435W,1997PASP..109..489T,2002ApJ...580L..39K,2006Natur.440..505L,2007MNRAS.381..201M,
2007ApJ...671.1944M,2010ApJ...713.1363C,2014MNRAS.442....2K,2016MNRAS.456.3157K,2017ApJ...834..118M}.

The maximum light in Type IIn SNe is primarily powered by an interaction of the ejecta with a massive CSM, implying 
that the major source of intrinsic polarization at maximum is due to this interaction, particularly with an asymmetric 
CSM. The luminosity from the interaction becomes progressively weaker as the CSM around the SN evolves into an optically 
thin state. The asymmetry in the SN photosphere (resulting from an aspherical explosion) can also complement the 
polarization signal arising from the CSM. The percentage contribution from the photosphere increases temporally and
plays a dominant role during the later phases as the luminosity from interaction dies off slowly. Overall, the relative 
contribution to the net luminosity from the photosphere and the CSM interaction decides the strength of net intrinsic 
polarization.

The high degree of intrinsic polarization at initial phases displayed by SN~2017hcc is an exceptional example among 
CCSNe. A significant decline of $\sim$\,3.5 per cent in the intrinsic polarization was seen within a period of $\sim$\,2
months. During the late phase ($\sim$\,2 months from maximum) polarization values are wavelength independent
(c.f. $\sim$\,1.4 per cent polarization in $V$ and $R$-bands). The decline in polarization seen in \hcc\ is very similar 
to the early decline observed in SN~2009ip \citep{2014MNRAS.442.1166M}, and explained as due to interaction with 
an aspherical CSM.
As discussed in Section~\ref{prog}, the progenitor of SN~2017hcc is likely an LBV star. These type of stars 
undergo episodic mass loss or giant eruption immediately before the explosion leading to the formation of 
circumstellar shells and/or equatorial disks/rings around them \citep{2017hsn..book..403S}. 
These structures distort a spherically symmetric SN photosphere upon interaction, enhancing the advancing 
polarization signal. While a detailed modelling is beyond the scope of this work, it appears that the strong 
polarization seen in SN~2017hcc could be a possible outcome of an interaction of the SN ejecta with an highly 
aspheric CSM \citep{2017ATel10911....1M}. 

Additionally, one can not rule out other possibilities, such as net reduction in the number density of the 
scatterers (such as electrons and dust grains) as a function of time in a non-spherical ejecta. This is 
expected because the cooling in the SN ejecta as a function of time (c.f. star symbols in Fig.~\ref{BB}) 
would result in a corresponding decay in the electron number density. Furthermore, the continuous expansion 
of SN ejecta as a function of time (c.f. blue filled circles in Fig.~\ref{BB}) may also cause a net reduction 
in the clumpiness (if any) leading to a decrease in the net polarization. 

\section{Summary}\label{sum}

We have presented imaging ({\textit Swift} UVOT), spectroscopy and polarimetry of the luminous Type IIn SN~2017hcc 
and compared the results with similar events. The light curve evolution was very slow during post-maximum phase 
($\sim$\,0.2 mag 100 d$^{-1}$ in $b$-band during $\sim$\,0\,--\,45 d). The $uvw1$, $uvw2$ and $uvm2$ band light curves 
evolved with similar rate however, the evolution was almost flat in $b$ and $v$ bands. The broad-band polarization 
characteristics were examined following robustly determined ISP contribution, which is found to be 0.17 per cent and 
140$^\circ$ (mainly arising from the Galactic dust and assuming a mean polarization efficiency). The intrinsic degree 
of polarization (in $V$-band) near the maximum light was $\sim$\,4.84 per cent which reduced to $\sim$\,1.36 per cent 
after $\sim$\,2 months. Similarly in $R$-band also, the intrinsic polarization values are high i.e. 3.22 per cent 
($\sim$\,+13 d) and 1.35 per cent ($\sim$\,+56 d). Such large amount of polarization is unusual for a SN. Interestingly, 
the intrinsic polarization angle was found to remain constant around $\sim$\,100\degr\ at all epochs. There is no 
significant change in the intrinsic polarization parameters even if maximum polarization efficiency relation is 
considered towards the SN direction.

We note that SN~2017hcc is not only peculiar in terms of polarization inferred but also belongs to the group of 
bright Type IIn events \citep[$M_{V}$ = --20.7 mag,][]{2017RNAAS...1a..28P}. Its peak luminosity is equivalent to 
SN~2006tf \citep[$M_{V}$ = --20.8 mag,][]{2008ApJ...686..467S}. The $\sim$\,27 d post-maximum spectrum of SN~2017hcc 
exhibited prominent narrow lines of H and He as typically seen in Type IIn SNe but the spectrum is bluer relative 
to SN~2010jl and SN~2012ab. The evolution of blackbody temperature and radius signifies a hotter and optically thick 
environment of SN~2017hcc than those of typical Type IIn SNe. The Na\,{\sc i} D lines are not visible in our spectrum, 
indicating a negligible host extinction. The mass-loss rate of SN~2017hcc is estimated to be 0.12 M$_{\sun}$ yr$^{-1}$ 
based on a simple CSM-ejecta interaction model which suggests an LBV progenitor for this event.

The high degree of polarization values found in SN~2017hcc support the fact that events belonging to this class are 
in general significantly polarized in comparison to other CCSNe and thermonuclear SNe (Type Ia) \citep{2008ARA&A..46..433W,
2017hsn..book.1017P}. A disc-like or toroidal CSM around the ejecta seems to be a common characteristic among Type 
IIn SNe that induces polarization signal. As described in Section~\ref{int_pol}, diversity exists in the polarization 
properties of Type IIn SNe. This indicates variation in the electron-scattering atmosphere which is regulated by 
complexity of the SN ejecta and/or CSM composition. A cumulative effect of the highly aspherical CSM and the ejecta 
density profile may have produced such a strong polarization in SN~2017hcc without significant change in the overall 
geometry \citep{2003ApJ...593..788K,2006ApJ...651..366K,2010ApJ...720.1500H,2017ApJ...837..105T}.
Better temporal coverage in polarimetric monitoring could reveal a wealth of crucial information including the 
CSM/ejecta geometry, clumpy structure and underlying progenitor nature etc. In this context the upcoming large 
aperture observing facilities shall be extremely useful. 

\section*{Acknowledgments}
We thank the referee for his/her several useful comments that have significantly improved the manuscript.
We also thank Masaomi Tanaka for useful discussions on various polarimetric aspects and Ashish Raj who kindly    
agreed to obtain HFOSC spectra during his observations. BK acknowledges the Science and Engineering Research 
Board (SERB) under the Department of Science \& Technology, Govt. of India, for financial assistance in the 
form of National Post-Doctoral Fellowship (Ref. no. PDF/2016/001563).
CE acknowledges (i) The National Key R \& D Program of China grant No. 2017YFA0402600 and
(ii) The International Partnership Program of Chinese Academy of Sciences grant No. 114A11KYSB20160008.
BK, DKS and GCA also acknowledge BRICS grant DST/IMRCD/BRICS/PilotCall1/MuMeSTU/2017(G) for the present work.
DKS and GCA also acknowledge DST/JSPS grant, DST/INT/JSPS/P/281/2018.
SBP acknowledges BRICS grant DST/IMRCD/BRICS/Pilotcall/ProFCheap/2017(G) for this work.

We thank the staff of IAO, Hanle and CREST, Hosakote that made HFOSC observations possible.
The facilities at IAO and CREST are operated by the Indian Institute of Astrophysics, Bangalore.
This work has made use of data from the European Space Agency (ESA) mission {\it Gaia}
(\url{https://www.cosmos.esa.int/gaia}), processed by the {\it Gaia} Data Processing and Analysis
Consortium (DPAC, \url{https://www.cosmos.esa.int/web/gaia/dpac/consortium}). Funding for the DPAC
has been provided by national institutions, in particular the institutions participating in the {\it Gaia}
Multilateral Agreement.
This research has made use of the SIMBAD data base, operated at CDS, Strasbourg, France.
This research made use of \textsc{RedPipe}\footnote{\url{https://github.com/sPaMFouR/RedPipe}},
an assemblage of data reduction and analysis scripts written by AS.

\bibliographystyle{mnras} 
\bibliography{sn17hcc}
\label{lastpage}
\end{document}